\documentclass{aa}
\usepackage{epsfig,graphicx,times}
\newcommand{\kms}{km~s$^{-1}$}
\newcommand{\teff}{{$T_\mathrm{eff}$}}
\newcommand{\logg}{{log~{\em g}}}

\newcommand{\vt}{{$\xi_\mathrm{t}$}}

\begin{document}
\title{B-type Supergiants in the SMC: Chemical compositions and comparison
of static and unified models}

%\subtitle{}

\author{
P.L. Dufton\inst{1}
\and
R.S.I. Ryans\inst{1}
\and 
C. Trundle\inst{2}
\and 
D.J. Lennon\inst{3}
\and
I. Hubeny\inst{4}
\and
T. Lanz\inst{5}
\and
C. Allende Prieto\inst{6}} 

\institute{Department of Physics and Astronomy, The Queen's University 
           of Belfast, BT7 1NN, Northern Ireland, UK       
     \and
       Insitituto de Astrof\'{i}sica, C/ V\'{i}a L\'{a}ctea s/n, E-38200,
       La Laguna, Tenerife, Spain
     \and
       Isaac Newton Group of Telescopes, Apartado de Correos 368, E-38700, 
       Santa Cruz de La Palma, Canary Islands, Spain
     \and 
       Steward Observatory, University of Arizona, Tucson, AZ 85712, USA 
     \and
       Department of Astronomy, University of Maryland, College Park,
       MD 20742, USA
     \and
       Department of Astronomy, The University of Texas at Austin, 
       Austin, TX 78712-1083, USA
}
\offprints{P.L. Dufton,
\email{p.dufton@qub.ac.uk}}

\date{Received  / Accepted}

\abstract{  
High-resolution UCLES/AAT spectra are presented for nine B-type supergiants 
in the SMC, chosen on the basis that they may show varying amounts of
nuclear-synthetically processed material mixed to their surface. These 
spectra have been analysed using a new grid of approximately 12\,000 
non-LTE line blanketed {\sc tlusty} model atmospheres to estimate 
atmospheric parameters and chemical composition. The abundance estimates
for O, Mg and Si are in excellent agreement with those deduced from
other studies, whilst the low estimate for C may reflect the use of the
\ion{C}{ii} doublet at 4267\AA. The N estimates are approximately an order
of magnitude greater than those found in unevolved B-type stars or 
\ion{H}{ii} regions but are consistent with the other estimates in 
AB-type supergiants. These results have been combined with results 
from a unified model atmosphere analysis of UVES/VLT spectra of B-type
supergiants (Trundle et al. \cite{Tru04}) to discuss 
the evolutionary status of  these objects. For two stars that are in common 
with those discussed by Trundle et al., we have undertaken a careful comparison
in order to try to understand the relative importance of the different 
uncertainties present in such analyses, including observational errors and 
the use of static or unified models. We find that even for these relatively 
luminous supergiants, {\sc tlusty} models yield atmospheric parameters 
and chemical compositions similar to those deduced from the unified code 
{\sc fastwind}.

\vskip 1.2truecm

\keywords{galaxies: Magellanic Clouds -- stars: abundances -- stars: early-type
-- stars: supergiants} 
}

\titlerunning{B-type Supergiants in the SMC}

\maketitle

%---------------------------------------------------------- SEC: INTRO ----
\section{Introduction}

The spectra of early-type supergiants provide an excellent method for studying
Local Group galaxies due to their high intrinsic luminosities (typically 1\,000
to 10\,000 L$_{\odot}$). Yet it is this extreme luminosity which also gives
rise to the complex nature of supergiants, whose stellar wind and extended
atmospheres drive them away from a plane-parallel, Local Thermodynamic
Equilibrium (LTE) regime. Developments in theoretical techniques that
incorporate non-LTE effects, line blanketing, sphericity  and mass-loss have
lead  to a resurgance of interest in these objects 
(for example see Hubeny \cite{Hub88};
Hubeny et al. \cite{Hub98};  Santolaya-Rey et al \cite{San97}; Hillier \& Miller
\cite{Hil98}). Recent studies of OB-type Supergiants  in the Local Group have
had a number of aims including understanding the evolution of massive stars,
studying the chemical composition of the host galaxy and calibrating the wind
momentum versus luminosity relation for distance determinations.  These studies
include those of Lennon et al. (\cite{Len91}), Gies \& Lambert (\cite{Gie92}),
Kudritzki et al. (\cite{Kud99}), McErlean et al. (\cite{McE99}) and
Repolust et al. (\cite{Rep04}) in our Galaxy, Fitzpatrick \& Bohannan
(\cite{Fit93}), Puls et al. (\cite{Pul96}), Dufton et al. (\cite{Duf00}), 
Korn et al. (\cite{Kor02}), Trundle et al. (\cite{Tru04, Tru05}) 
and Venn (\cite{Ven99}) for the Magellanic System, 
Venn et al. (\cite{Ven00}), Smartt et al. (\cite{Sma01}) and Trundle  et al.
(\cite{Tru02}) for M31, Monteverde et al. (\cite{Mon97, Mon00}) and   Urbaneja
et al. (\cite{Urb02}) for M33 and Kaufer et al. (\cite{Kau04}),  Urbaneja et
al. (\cite{Urb03}) and Venn et al. (\cite{Ven01, Ven03b})  for other Local
Group galaxies. Studies of the progenitor O-type stars (see Bouret et al.
\cite{Bou03}, Hillier et al. \cite{Hil03}, Heap et al. \cite{Hea04}) 
and B-type giants (see Korn et al. 
\cite{Kor02}, Lennon et al. \cite{Len03}) have also provided relevant
observations to aid our understanding of the evolution of the surface chemical
composition of early-type stars.

Since the studies of Jaschek \& Jaschek (\cite{Jas67}), Walborn (\cite{Wal72})
and Dufton (\cite{Duf72}), the dispersion of observed nitrogen abundances in
OBA-type stars selected within a particular metallicity environment has
stimulated continuing interest in the area of massive stars. These changes in 
the surface chemical composition of hot stars are closely coupled to their 
rotational velocities. Current theoretical models require initially high 
rotational velocities in order to mix nucleosynthetically processed material 
to the surface (Heger \& Langer \cite{Heg00}; Maeder \& Meynet \cite{Mae00, 
Mae01}). In turn these models predict that B-type supergiants and giants 
should still be rotating relatively rapidly with larger rotational velocities 
in lower metallicity regions where the loss of angular momentum through 
the strong stellar wind is less significant. 
However as discussed by Howarth et al. (\cite{How94}), Lennon et al.
(\cite{Len03}) and Trundle et al. (\cite{Tru04}), this appears inconsistent
with the observed widths of the metal line spectra. Ryans et al. (\cite{Rya02})
have attempted to distinguish between the different broadening mechanisms (viz.
microturbulence, rotation and macroturbulence). For a sample of Galactic
supergiants, they found that the contribution of rotation to the
line-broadening was relatively small, in turn leading to large discrepancies
with the theoretical predictions. 

In this paper, we present high resolution observations of a sample of SMC
supergiants chosen from a preliminary exploration of B-type supergiants by
Dufton et al. (\cite{Duf00}). The selection of targets was made on the basis
that they might show different degrees of nucleosynthetic processed material at
their surface. These have been analysed using a new grid of approximately
12\,000  models computed with the non-LTE codes {\sc tlusty} 
and {\sc synspec} (Hubeny 
\cite{Hub88}; Hubeny \& Lanz \cite{Hub95}; Hubeny et al. \cite{Hub98}). These
models do not include any contribution from a stellar wind and therefore, for
two targets, we have compared our results with those deduced
by Trundle et al. (2004) from the unified code {\sc fastwind} 
(Santolaya-Rey et al. \cite{San97}; with some updates 
described in Herrero et al. \cite{Her02} and Repolust et al. 
\cite{Rep04}) to investigate the
magnitude of both observational and theoretical  uncertainties. In a companion
paper, we intend to obtain reliable  estimates of the actual projected
rotational velocities of our own and  other SMC targets using the methods
discussed by Ryans et al. (\cite{Rya02}). These should allow a better
understanding of the relationship between the chemical evolution and rotation
for massive stars.

%------------------------------------------------------------ tab: obs --------
\begin{table}
\caption{Observational summary of the nine program stars. Spectral types are
those assigned by Lennon (1997). Apparent  V-magnitudes are taken from Massey 
(2002) where available and supplemented by Garmany et al. (1987;$^\dagger$)}
\label{tab:obs}
%\centering
\begin{tabular}{llllrr}
\hline
\noalign{\smallskip}
 Star     &     V  & Spec.      &  S/N ratio  &$ v_{\mathrm r}$ & FWHM \\
          &  [mag] & Type       &             & \multicolumn{2}{c}{\kms}   \\
\noalign{\smallskip}
\hline\noalign{\smallskip}
AV\,78    & 11.05          & B1.5Ia$^+$  & 130 & 178 & 80  \\  
AV\,215   & 12.69          & BN0Ia       &  90 & 159 & 115 \\  
AV\,242   & 12.11$^\dagger$ & B1Ia        & 110 & 142 & 100 \\  
AV\,303   & 12.78          & B1.5Iab     & 100 & 186 & 65  \\  
AV\,374   & 13.09          & B2Ib        &  65 & 125 & 70  \\  
AV\,462   & 12.54          & B1.5Ia      & 115 & 126 & 70  \\  
AV\,472   & 12.62          & B2Ia        &  90 & 138 & 70  \\  
AV\,487   & 12.58          & BC0Ia       & 125 & 183 & 100 \\  
SK\,191   & 11.86$^\dagger$ & B1.5Ia      & 125 & 134 & 130 \\  

\noalign{\smallskip}\hline
\end{tabular}\\
\smallskip
\end{table}
%-----------------------------------------------------------------------------

%---------------------------------------------------------- fig: smcw ---------
%\begin{figure}
%\hspace*{-4mm}\psfig{file=magellanic.ps,width=9cm}
%\caption{Schematic diagram of the LMC, SMC and the Inter-Cloud Region. Solid
%lines define the stellar concentrations within the Magellanic Clouds, while the
%dashed lines show the H~{\sc i} envelopes of the Magellanic system. The
%positions of our targets within the SMC wing are marked by solid black
%circles.} 
%\end{figure}
%-----------------------------------------------------------------------------

%----------------------------------------------------------- SEC: OBS ---------
\section{Observations and data reduction}
\label{obs}

The high-resolution, spectroscopic data presented here were obtained during an
observing run with the 3.9-m Anglo-Australian Telescope (AAT) from
29 September to 1st October 1998 inclusively. The University College of London \'{E}chelle Spectrograph (UCLES) was
used with the 31 lines mm$^{-1}$ grating and with a TeK 1K$\times$1K CCD, 
providing complete spectral coverage between $\lambda\lambda$3900--4900~\AA\/ 
at a FWHM resolution of $\sim$0.1~\AA. Conditions were excellent throughout the
three night run, with  stellar exposures being bracketed with Cu-Ar  arc
exposures for wavelength calibration. Observations were obtained for  nine
supergiants and these are summarized in Table~\ref{tab:obs}. Listed are the
stellar V-magnitude (taken from Garmany et al. \cite{Gar87} and Massey
\cite{Mas02}), spectral types (from Lennon  \cite{Len97}), the signal-to-noise
(S/N) ratio at approximately 4500\AA, the  heliocentric radial velocity
($v_{\mathrm r}$) and the typical full-width-half-maxima (FWHM) 
of the metal absorption
line spectrum. The latter two quantities  were estimated as discussed below,
whilst the S/N ratios were deduced by fitting low order polynomials to the
continuum. Care was taken to include continuum regions from different parts of
the blaze profile  and hence these estimates should be considered as
conservative, particularly given the significant wavelength overlap between
adjacent orders.

The choice of targets was based on the spectral types deduced by Lennon
(\cite{Len97}) and the preliminary non-LTE analysis of Dufton et al.
(\cite{Duf00}) of intermediate dispersion spectroscopy. In particular, we
attemped to sample a range of luminosity types (ranging from Ia$^+$\, to Ib)
and with varying degrees of mixing of nucleosynthetic material to the surface.
For the latter, we included both BN and BC spectral types and stars with
different ratios of \ion{N}{ii} to \ion{C}{ii} equivalent widths (see Fig.\,9
of Dufton et al.).

The two dimensional CCD datasets were reduced using standard procedures  within
IRAF\footnote{IRAF is written and supported by the {\sc
IRAF} programming group at the National Optical Astronomy Observatories (NOAO)
in Tucson (http://iraf.noao.edu)} (Tody \cite{Tod86}). Preliminary processing of the CCD frames 
such as over-scan correction, trimming of the data section and flat-fielding 
were performed using the {\sc CCDRED} package (Massey \cite{Mas97}), whilst 
cosmic-ray  removal, extraction of the stellar spectra, sky-subtraction and
wavelength  calibration were carried out using the {\sc SPECRED} (Massey et al.
\cite{Mas92})  and {\sc DOECSLIT} (Willmarth \& Barnes \cite{Wil94})
packages.

Equivalent widths (EWs) of lines were measured using the STARLINK spectrum 
analysis program DIPSO (Howarth et al.\ \cite{How94}). Low order polynomials 
were fitted to the adjacent continuum regions for normalisation and 
Gaussian profiles to the metal and non-diffuse helium lines using  non-linear
least square routines. As discussed by Ryans et al. (\cite{Rya03}),  the
profiles of metal absorption lines in the spectra of early-type  supergiants
are well represented by a Gaussian profiles. Tests using a different profile
shape showed that this assumption was not critical to the fitting procedure.

In addition the central wavelength and widths of the metal lines   could be
used to determine the stellar radial velocity and   FWHM, which are summarized
in Table~\ref{tab:obs}. Typically 10 relatively  strong, isolated lines were
used for  the former, whilst the FWHM were estimated from  the \ion{Si}{iii}
multiplet near 4560\AA, which was well observed in all  our spectra. Standard
deviations were typically $\pm$2 \kms and  $\pm$5 \kms respectively, with the
values for the FWHM being rounded to the nearest 5 \kms. For the hydrogen
lines, the  identification of the continuum was complicated by the presence of
a significant echelle ripple. In these cases the continua of the adjacent 
orders were  moved to a common wavelength scale, fitted using low order 
polynomials and merged. This merged spectrum, which should represent
the echelle ripple for the order containing the hydrogen line was then 
used to rectify the relevant order prior to normalisation.
Profiles  were then extracted with the continuum levels being defined at
$\pm$16 \AA\/  from the line centre.

%--------------------------------------------------------- SEC: model ------- 
\section{Spectral analysis}
\label{analysis}
\subsection{Model atmosphere calculations} \label{model}

The analysis is based on grids of non-LTE model atmospheres calculated 
using the codes {\sc tlusty} and {\sc synspec} (Hubeny \cite{Hub88}; Hubeny \& Lanz 
\cite{Hub95}; Hubeny et al. \cite{Hub98}). Normally when using such codes 
to analyse an observational dataset, the approach is  to undertake 
specific calculations, often in the form of grids covering a restricted 
range of atmospheric parameters. Such an approach has
the advantage that the calculations are tailored to the specific problem.
Additionally, non-LTE codes often exhibit stability problems, and these are
particularly severe when trace ionic species (e.g. Si~{\sc ii} at higher
effective temperatures or Si~{\sc iv} at lower effective temperature) are 
included. By limiting the range of atmospheric parameters considered, and
by selecting the model ions to be included, these stability problems can be
ameliorated. However this approach is inefficient when attempting
to analyse either large observational datasets or different datasets
for similar types of objects. 

We are therefore developing an approach based on the calculation of large
grids covering the range of atmospheric parameters appropriate to B-type
stars. One fundamental assumption is that the metal line blanketing is 
dominated by Fe and we return to its validity later
in this section. Then the structure of the atmosphere is defined by the 
Fe abundance (or `metallicity'), the effective temperature (\teff), 
gravity ($g$) and microturbulence (\vt, which affects the
amount of blanketing). Currently we are calculating 4 grids with
metallicities appropriate to our Galaxy (i.e. $[\frac{Fe}{H}]$ = 7.5 dex),
the LMC (metallicity reduced by 0.3 dex), SMC (metallicity reduced 
by 0.6 dex), and low metallicity regimes (metallicity reduced by 1.1 dex).
Due to limitations on the available CPU power we adoped the 
`classical' {\sc tlusty} ODF Fe model ions for these calculations. 
The models ions for the light elements that have been explicitly 
included in the non-LTE
calculations, together with their origins and from where they may be
obtained, are summarized in Table \ref{tab:ions}. 
These models come from two main sources, which are discussed in, eg.
Allende Prieto et al. (\cite{All03}) and Lanz and Hubeny (\cite{Lan03}).

\begin{table}
\caption{Source of model ions used in generating the non-LTE model
atmosphere grids. These model ions are  discussed in Lanz and Hubeny 
(\cite{Lan03}; designated {\sc tlusty}) and Allende Prieto et al. 
(\cite{All03}; designated Allende). They are available from 
{\it http://tlusty.gsfc.nasa.gov/} and {\it http://hebe.as.utexas.edu/} 
respectively.
} 
\label{tab:ions}
\begin{center} 
\begin{tabular}{lcc} 
\hline \noalign{\smallskip}

Ion             & Levels  & Source \\
\ion{H}{i}      & 16      & {\sc tlusty} \\
\ion{He}{i}     & 24      & {\sc tlusty} \\
\ion{He}{ii}    & 14      & {\sc tlusty} \\
\ion{C}{ii}     & 39      & {\sc tlusty} \\
\ion{C}{iii}    & 23      & {\sc tlusty} \\
\ion{N}{ii}     & 51      & Allende \\
\ion{N}{iii}    & 32      & {\sc tlusty} \\
\ion{O}{ii}     & 74      & Allende \\
\ion{O}{iii}    & 29      & {\sc tlusty} \\
\ion{Mg}{ii}    & 31      & Allende \\
\ion{Si}{ii}    & 46      & Allende \\
\ion{Si}{iii}   & 74      & Allende \\
\ion{Si}{iv}    & 23      & {\sc tlusty} \\
\ion{S}{ii}     & 14      & {\sc tlusty} \\
\ion{S}{iii}    & 20      & {\sc tlusty} \\
\ion{Fe}{ii}    & 35      & {\sc tlusty} \\
\ion{Fe}{iii}   & 50      & {\sc tlusty} \\
\ion{Fe}{iv}    & 43      & {\sc tlusty} \\

\noalign{\smallskip}\hline
\end{tabular}\\
\end{center}
\smallskip
\end{table}

\begin{figure}[ht]
\includegraphics[angle=0,width=3.4in]{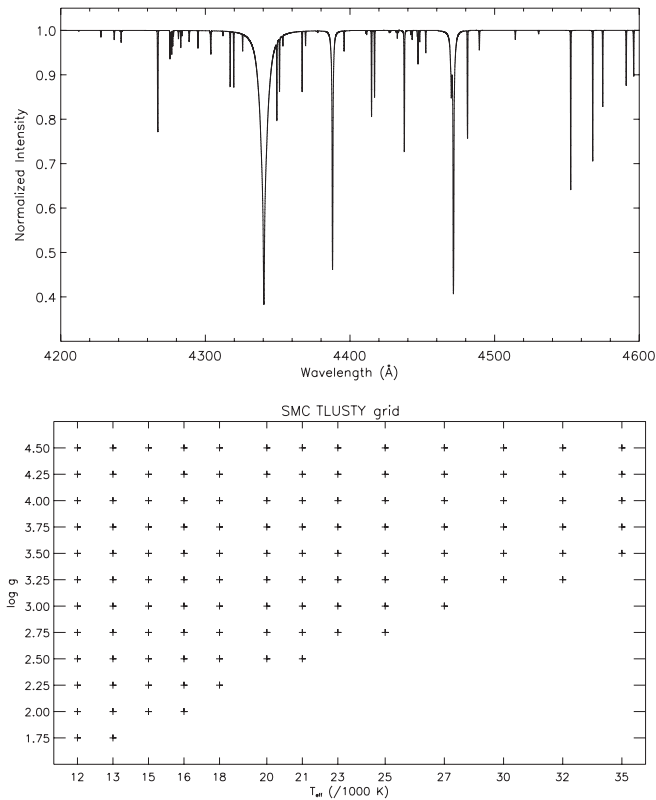}
\caption{
The bottom panel indicates the range of effective temperatures (\teff) and
gravities (\logg) covered by our proposed grids, with the
lowest gravity models being close to the Eddington limit. 
At each point, 100 models are calculated covering a
range of microturbulence, metallicity and light metal abundances. In total,
this leads to approximately 12\,000 models. 
The top panel shows the predicted spectrum  for a small wavelength region
from a typical model with an 
effective temperature of 20\,000K, logarthmic gravity of 3.0 dex,
microturbulence of 10 \kms and default light element abundances for the 
SMC.}
\label{grid}
\end{figure}

For each metallicity (i.e. Fe abundance), we consider approximately 
120 different \teff-\logg\ 
points (see Fig. \ref{grid}), which are appropriate to B-type stars 
ranging from the main-sequence to the Eddington limit. At each 
\teff--\logg\ point, we then calculate 25 models covering 5 values 
of \vt\, and 5 sets of {\it light} metal
abundances, which are appropriate to the underlying metallicity regime.
This corresponds to approximately 3\,000 models per metallicity or 12\,000
models in total. 

The analysis of a spectrum is then relatively straightforward 
and fast. One selects an appropriate metallicity grid, and then analyses 
spectral features to determine first the atmospheric parameters and
then the stellar chemical composition. The process is currently 
semi-automated, and will be further automated in the light of experience
gained regarding those areas which will continue to require user intervention.
Further information (on for example the spectral
lines incorporated in synthesis calculations and the light element abundances
adopted in different grids) is available at http://star.pst.qub.ac.uk/.

The major difficulty with the approach outlined above is to maintain stability
over the wide range of atmospheric parameters (illustrated in Fig. \ref{grid}),
which in turn leads to different ionic species and indeed atomic processes
being important in different parts of the grid. Closely associated with
this has been the problem of quality control for the very large number of
proposed models. However we have now completed the SMC metallicity grid 
and have tested this against existing calculations using {\sc tlusty}
and {\sc detail/surface} (see Becker \& Butler \cite{Bec88,Bec89,Bec90};
Butler \cite{But84}; Giddings \cite{Gid81}; 
Korn et al. \cite{Kor02}; McErlean et al. \cite{McE99} for details 
of atomic data and codes). Both the overall agreement and expected differences 
(due e.g. to improved model ions) are highly encouraging. Difficulties
remain for a small number of spectral features (see http://star.pst.qub.ac.uk/
for details), but these are relatively minor and should not affect 
the usefulness of the grids. 

These models have been used to calculate spectra (see Fig. \ref{grid}), which 
in turn provide theoretical hydrogen and helium line profiles and equivalent 
widths for light metals for a range of abundances. Note that as we keep the 
iron abundance fixed within any given grid, we have less extensive data for 
this element. Tests showed that the grid spacings for our
atmospheric parameters was sufficiently fine to allow reliable interpolation 
of the theoretical hydrogen and helium profiles at intermediate values. For 
the theoretical metal line equivalent widths, these can then be accessed 
via a GUI interface written in IDL, which allows the user to interpolate 
in order to calculate equivalent widths and/or abundance estimates for 
approximately 200 metal lines for any given set of atmospheric 
parameters. Ryans et al. (\cite{Rya03}) reported that the increment of 0.4 dex 
used in our grids was fine enough to ensure that no significant errors were 
introduced by the interpolation procedures. Full theoretical spectra are also 
available for any given model. In summary these grids allow a user-friendly 
non-LTE analysis of hydrogen and helium line profiles and of the profiles and 
equivalent width of the lines of light elements over a range of iron line 
blanketing and atmospheric parameters appropriate to B-type stars in our Galaxy 
and in Local Group galaxies such as the Magellanic Clouds.

\begin{figure}[ht]
\includegraphics[angle=0,width=3.4in]{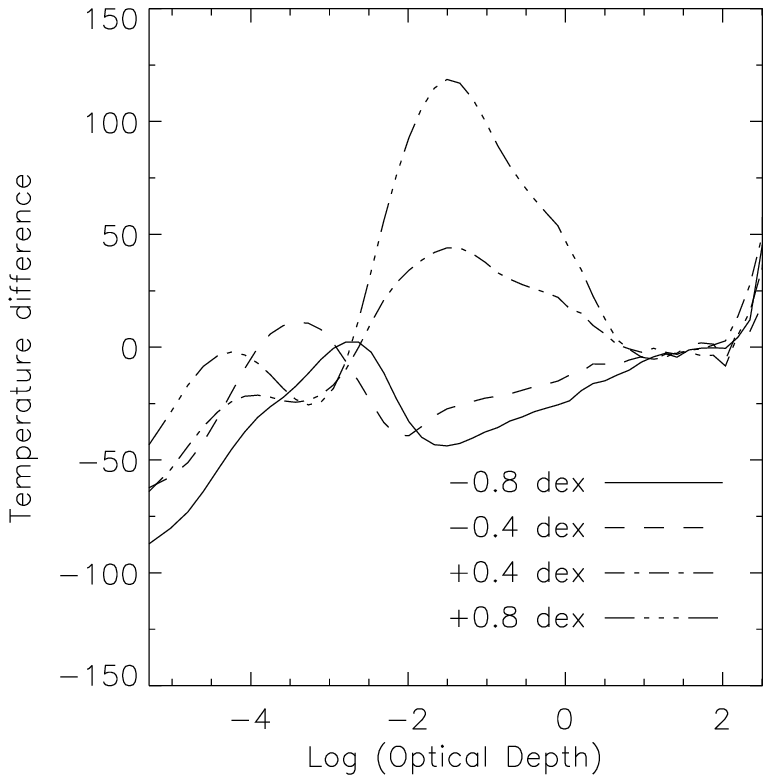}
\caption{
Differences in the temperature scales as a function of the Rosseland mean 
optical depth for models with an effective
temperature of 20\,000\,K, logarithmic gravity of 2.5 dex, microturbulence
of 10\,\kms\, and an Fe abundance appropriate to the SMC. There are five
different models with baseline SMC element abundances and with light element
abundances increased and decreased by 0.4 and 0.8 dex. The four plots show the
difference in the temperature structure of the latter compared with 
that of the baseline abundance model. The small differences imply
that the temperature structures are effectively independent of the
light element abundances adopted.}
\label{t_structures}
\end{figure}

This approach is based on several assumptions. Firstly there are those implicit
in the use of the {\sc tlusty/synspec} package. Particularly relevant to the 
current analysis are the assumptions of a static atmosphere (which precludes the
inclusion of a wind) and of a plane parallel geometry. We will return to these
assumptions when comparing the results with those generated using the unified
code {\sc fastwind} (Santolaya-Rey et al. \cite{San97}; Herrero et al.
\cite{Her02}). Secondly although we have included line blanketing due
to Fe, we have excluded that due to other iron peak elements, due to limits on 
the available computational power. However experience has shown that the bulk 
of the opacity in the regime of interest is due to Fe (Hubeny et al.
\cite{Hub98}) so this simplification is unlikely to be a source of significant 
error. Thirdly we have also implicitly assumed that the atmospheric structure
is not affected by the adopted light element (such as CNO, Mg, Si) abundances, 
which are constrained to vary in step. To test this assumption we have  
considered representative atmospheric parameters spanning our ranges of 
effective temperature, gravity, microturbulence and iron abundance. For each
set of atmospheric parameters, we have then fixed,
for example, the O abundance but recalculated models allowing the other light
element abundances to vary. If the assumption that the atmospheric structure
does not depend on the light element abundances is correct, the oxygen
spectrum should not vary significantly within these calculations. Our 
tests show that, excluding very weak lines, variations in equivalent width 
are typically less than 1\% and are always less than 5\%. Additionally
we have checked that our temperature structures for given sets of atmospheric 
parameters (and iron abundance) do not depend significantly on the light 
element abundances. This is illustrated in Fig. \ref{t_structures}, where it
can be seen that the temperature structures agree to typically $\pm$50K
with the maximum differences being less than $\pm$150K. Hence we 
believe that the assumption that the atmospheric structure depends to first
order on the atmospheric parameters and iron abundance is acceptable. 

%--------------------------------------------------------- tab: atm_stars ------- 
\begin{table*}
\caption{Atmospheric parameters and abundances of program stars. Also
listed are results for the main sequence SMC target, AV\,304. The number
of lines considered is shown in brackets.}
\label{tab:atm_stars}
\begin{tabular}{lllllllllllll}
\hline\hline\noalign{\smallskip}
& AV\,78 & AV\,215 & AV\,242 & AV\,303 & AV\,374 & AV\,462 & AV\,472 & AV\,487 & SK\,191
& AV\,304
\\
\hline\\																													
\teff &	21250 & 26500 & 21500 & 18000 & 18500 & 19000 & 19000 & 27000 & 20500
& 27000
\\
log $g$	& 2.45 & 3.00 & 2.50 & 2.30 & 2.50 & 2.40 & 2.40 & 2.95 & 2.45
& 3.9 
\\
\vt & 13 & 13 & 13 & 14	& 14 & 16 & 15 & 11 & 16 
& 3
\\
\hline
\\
C        & 6.91 (1)  & -- & 7.06 (1) & 7.15 (3) & 7.07 (1) 
& 7.00 (3) & 7.06 (3) & 7.24 (1) & 7.00 (1) & 7.36 (3) 
\\
$\sigma$  & -- & -- & -- & 0.09 & -- & 0.07 & 0.12 & --   & -- & --
\\
\\
N & 7.80 (9) & 7.60 (3) & 7.09 (3) & 7.28 (6) & 7.18 (5) & 7.42 (8) &  7.51 (9) 
& $<$7.3 & 7.53 (4) & 6.55 (1)
\\
$\sigma$ & 0.06 & 0.06 & 0.07 & 0.05 & 0.05 & 0.05 & 0.03 & -- & 0.02 & --		
\\
\\
O & 7.74 (12) &	 8.00 (11) & 8.11 (19) & 8.34 (22) & 8.26 (12) & 8.12 (17)
& 8.05 (18) & 8.09 (10) & 8.11 (13) & 8.13 (42)
\\
$\sigma$ & 0.08	& 0.08 & 0.04 & 0.03 & 0.03 & 0.05 & 0.03 & 0.04 & 0.06 & --
\\
\\
Mg & 6.79 (1) & 6.67 (1) & 6.60 (1) & 6.69 (1) & 6.62 (1) & 6.73 (1)
& 6.67 (1) & 6.79 (1) & 6.71 (1) & 6.77 (1)
\\
\\
Si & 6.90 (4) & 6.87 (4) & 6.65 (4) & 7.01 (4) & 7.01 (4) & 6.85 (6) & 6.81 (3) 
& 6.93 (3)& 6.64 (3) & 6.75 (6)
\\
$\sigma$ & 0.01 & 0.02 & 0.01 & 0.03 & 0.02 & 0.01 & 0.01 & 0.03 & 0.01 & --
\\
\noalign{\smallskip}   \hline
\noalign{\smallskip} \hline     
\end{tabular}
\end{table*}    
%------------------------------------------------------------------------------- 

\subsection{Estimation of atmospheric parameters}

The atmospheric parameters of our targets were estimated using standard
techniques, viz. the silicon ionization equilibrium for effective temperaure,
the profiles of the Balmer lines for gravity and the relative strength of
absorption lines of an ionic species for microturbulence. These techniques have
been described by, for example, Kilian (\cite{Kil92}),  McErlean et al.
(\cite{McE99}), Korn et al. ({\cite{Kor02}) and Trundle et al.  (\cite{Tru04})
and will only be briefly discussed here. However it should be noted that the
estimation of the atmospheric parameters is an iterative process as they are
inter-related. Additionally all the results presented below are based on the
grid with an iron abundance appropriate to the SMC (i.e. 0.6 dex less than
solar - Anders \& Grevesse \cite{And89}). Tests were undertaken using other
grids and these are discussed below.

\subsubsection{Effective temperature, \teff} \label{teff} Initial estimates of
the surface gravity were based on the calibration of McErlean et al.
(\cite{McE99}). These were then used to estimate the effective temperature
normally using the \ion{Si}{iii} to \ion{Si}{iv} ionization equilbrium. For
AV\,374, an additional estimate was available from the \ion{Si}{ii} to
\ion{Si}{iii} ionization equilbrium, with the two values agreeing to within
200\,K. The \ion{Si}{iii} multiplet at
approximately 4560\AA\  contains lines with a range of equivalent widths and
hence the microturbulence could be simultaneously constrained (see Sect.
\ref{vt}). The final effective temperature estimates are summarized in Table
\ref{tab:atm_stars}, rounded to the nearest 500\,K. The quality of the 
observational data implies that these
will have an uncertainty of typically $\pm$1\,000-2\,000\,K (see Sect.
\ref{fast_tlusty}), while the assumptions implicit to {\sc tlusty} will
contribute additional uncertainties. We note that Lee et al.\ (\cite{Lee04})
have found that using grids with different Fe abundances changed the estimates
by  typically 500K or less and hence should not be a major source of error.

\begin{figure*}[ht]
\includegraphics[angle=270,width=6.0in]{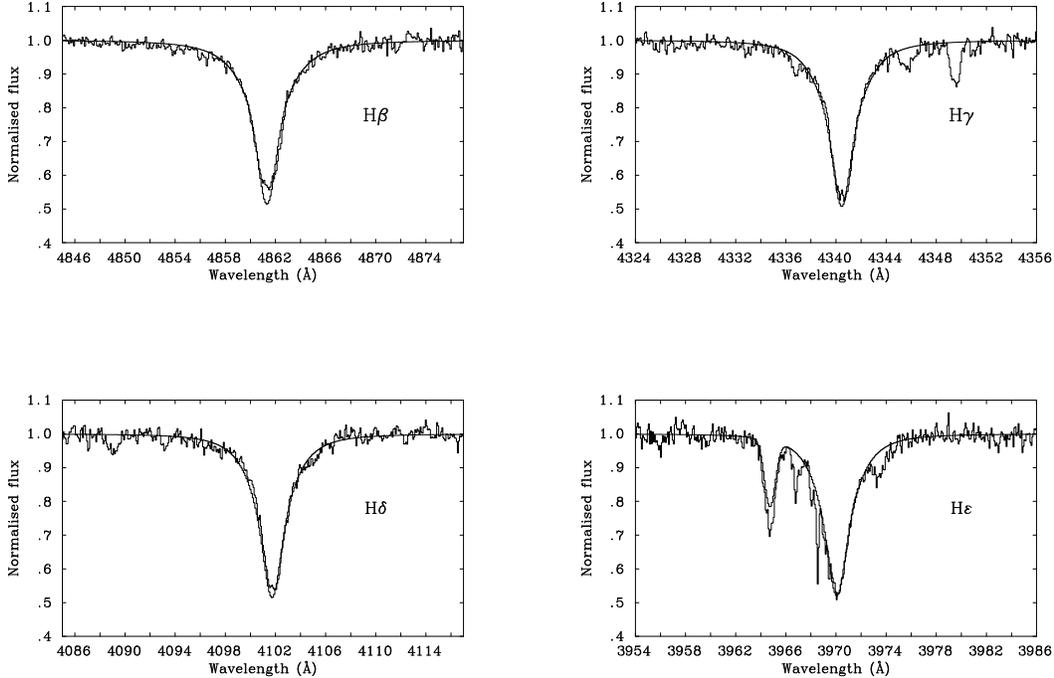}
\caption{Observed and theoretical Balmer line proflies for AV\,472. The 
theoretical profiles have been convolved with a Gaussian profile using 
the FWHM listed in Table \ref{tab:obs}. Note the filling in of the Balmer line
cores due to wind emission and that this is more significant in the lower 
series members.
}
\label{grav_det}
\end{figure*}

\subsubsection{Surface gravity, $g$} \label{grav}

The surface gravity was deduced by fitting theoretical profiles to the
Balmer series features, H$\beta$ to H$\epsilon$. As expected the lower
series members (and in particular H$\beta$) showed evidence of their 
profiles being affected by a stellar wind and in these cases more weight 
was given to the higher series members. The quality of the fit is shown
in Fig. \ref{grav_det} for the target AV\,472, where the theoretical profiles
have been convolved with a Gaussian profile 
to allow for rotational and macroturbulent broadening. Use of 
grids with different Fe abundances did not affect the gravity determinations,
with the major source of error (of the order of $\pm$0.2 dex) arising from
observational and fitting uncertainties. The final adopted gravities, $g$
(in units of cm s$^{-2}$) are listed in Table \ref{tab:atm_stars}).

\subsubsection{Microturbulence, \vt} \label{vt}

The microturbulent velocity was initially determined from the  \ion{Si}{iii}
multiplet at approximately 4560\AA, by minimising the spread  of abundances
deduced from its 3 members. Subsequently the same procedure  was undertaken
using the rich \ion{O}{ii} spectra in our targets. As has been  found previously
(see, for example, McErlean et al. \cite {McE99}, Vrancken et al. \cite{Vra00}
and Trundle et al. \cite{Tru04}), the \ion{O}{ii} spectra gave estimates which
were systematically  larger by between 5 and 10 \kms. Typically 10-20 
\ion{O}{ii} lines are included but the analysis is complicated by them arising
from different multiplets. Uncertainties in the atomic data and in the magnitude
of the non-LTE effects may affect such an analysis and hence the source of these
differences is unclear. We will follow previous authors in adopting the results
from the silicon multiplet but note that the estimates must be considered 
uncertain by at least 5 \kms (obviously this uncertainty will have a larger 
effect on the abundances derived from strong lines as the weak lines 
are unaffected by the microturbulence, as is apparent from 
Table~\ref{tab:atm_err}). Tests using grids with different base 
metallicities yielded, as would be expected, effectively identical 
estimates for the microturbulence and the adopted values are again listed 
in Table  \ref{tab:atm_stars}.

\subsection{Photospheric Abundances}  
\label{abund}

The adopted atmospheric parameters (listed in Table~\ref{tab:atm_stars})  were
used to derive absolute non-LTE abudances for  the programme stars. Tests were
undertaken by comparing the observed equivalent widths with theoretical values
from different metallicity  grids. The adopted iron abundance was not found to
significantly affect  the light element abundance estimates (as had also been
found by Lee et al. \cite{Lee04}) and the results presented in Table
\ref{tab:atm_stars} are  for the SMC grid (metallicity -0.6 dex compared with 
the solar estimate of Anders \& Grevesse \cite{And89}). For example, we 
re-analysed the metal lines equivalent widths for AV\,215 (one of the targets 
considered further in Sect. \ref{fast_tlusty}) using the grids of models with 
metallicities appropriate to the LMC (-0.3 dex) and low metallicity regimes 
(-1.1 dex). The average change in the light element abundance estimates 
was 0.04 dex, with all changes being less than 0.1 dex.

Also listed in this Table are error estimates  due to the
scatter in the abundances deduced from individual features (note that there
will be other sources of uncertainty including errors in the adopted
atmospheric parameters or in the physical models which we address below and in
Sect. \ref{fast_tlusty} respectively). These estimates assume that the errors
are normally distributed and are simply the sample standard deviations divided
by the square root of the number of spectral features included. For ions, such
as \ion{Mg}{ii}, where only one spectral feature was observed, the sample
standard deviations from other ions imply that an error estimate of 0.1 to 0.2
dex is probably appropriate. 

VLT observations of the SMC main sequence B-type star, AV\,304
(Rolleston et al. \cite{Rol03}) have been re-analysed with our {\sc tlusty}
grids (Hunter et al. \cite{Hun04}). This star is particularly suitable for
analysis as its spectrum has relatively sharp absorption lines. In  Table
\ref{tab:atm_stars}, we list the atmospheric parameter and
abundance estimates for AV\,304 and note that
these are in reasonable agreement with those deduced from the LTE analysis of
Rolleston et al. (\cite{Rol03}), particularly when the non-LTE corrections of
Lennon et al.\ (\cite{Len03}) are included. We believe that  these results
provide a baseline for the current chemical composition of the SMC and given
the methods used are particularly appropriate for comparison with our results
for the SMC supergiants. Hence we have undertaken a differential analysis for
each supergiant relative to AV\,304 and these results are presented in
Table~\ref{tab:diff}, together with error estimates calculated using the same
methodology as for the absolute abundances. We expect that these
differential abundances are more reliable as they would be less prone to
uncertainties in the atomic data.
Some indirect evidence for this is provided by the smaller error estimates that
are found for the differential, rather than the absolute abundance estimates.

%--------------------------------------------------------- tab: diff ------- 
\begin{table*}
\caption{Differential abundances of program stars relative to 
the main sequence SMC target, AV\,304. The number of lines considered
is shown in brackets.}
\label{tab:diff}
\begin{tabular}{lllllllllllll}
\hline\hline\noalign{\smallskip}
& AV\,78 & AV\,215 & AV\,242 & AV\,303 & AV\,374 & AV\,462 & AV\,472 & AV\,487 & SK\,191
\\
\hline
\\
C        & -0.38 (1)  & -- & -0.23 (1) & -0.13 (3) & -0.22 (1) 
& -0.36 (3) & -0.30 (3) & -0.05 (1) & -0.29 (1) 
\\
$\sigma$  & -- & -- & -- & 0.07 & -- & 0.07 & 0.10 & --   & -- 
\\
\\
N & 1.37 (1) & 1.00 (1) & 0.41 (1) & 0.79 (1) & 0.46 (1) & 0.93 (1) &  0.96 (1) 
& $<$0.7 (1) & 0.98 (1)
\\
\\
O & -0.43 (12) & -0.13 (11) & -0.03 (19) & 0.20 (22) & 0.10 (12) & 0.01 (17)
& -0.11 (18) & 0.00 (10) & 0.03 (13)
\\
$\sigma$ & 0.08	& 0.04 & 0.03 & 0.03 & 0.04 & 0.03 & 0.03 & 0.03 & 0.03
\\
\\
Mg & 0.02 (1) & -0.10 (1) & -0.17 (1) & -0.08 (1) & -0.15 (1) & -0.04 (1)
& -0.10 (1) & 0.02 (1) & -0.06 (1) 
\\
\\
Si & 0.12 (4) & 0.08 (4) & -0.14 (4) & 0.22 (4) & 0.21 (4) & 0.08 (4) & 0.04 (3) 
& 0.14 (3)& -0.15 (3)
\\
$\sigma(mean)$ & 0.05 & 0.05 & 0.05 & 0.08 & 0.07 & 0.04 & 0.04 & 0.05 & 0.06
\\
\noalign{\smallskip}   \hline
\noalign{\smallskip} \hline     
\end{tabular}
\end{table*}    
%------------------------------------------------------------------------------- 

There will be other potential sources of uncertainty in the abundance estimates
presented in Tables \ref{tab:atm_stars} and \ref{tab:diff}. For example errors
in the adopted atmospheric parameters would systematically affect the
estimates from any given ionic species. Therefore in Table \ref{tab:atm_err},
we list the changes in the abundance estimates that would arise from
increases in the values of \teff, \logg\  and \vt\  by 1000K, 0.2 dex and 5\kms\,
respectively. Note that these should be considered as indicative as they
may vary from line to line in any given ionic species. Values are listed for 
two stars, AV\,303 and AV\,487, which were chosen as they were the coolest
and hottest stars in our sample. It is encouraging to  note that given the
uncertainties in estimating the mictoturbulent velocities, the errors
associated with this quantity are relatively small. By contrast the
errors associated with the effective temperature and gravity are typically 
0.1--0.2 dex but can be as large as 0.3 dex. These estimates are appropriate 
to the absolute abundance estimates listed in Table \ref{tab:atm_stars}. In the
case of estimates listed in Table \ref{tab:diff}, systematic
errors in the derivation of the atmospheric parameters might lead to smaller
errors in the differential abundances and hence these estimates should be
considered as conservative.

%--------------------------------------------------------- tab: atm_err ------- 
\begin{table*}
\caption{Effects of changes in the adopted atmospheric parameters by
$\Delta$\teff = 1,000K, $\Delta$log g=0.2 and $\Delta$\vt=5\kms. Results
are presented for the coolest (AV\,303) and hottest (AV\,487) stars
in our sample.}
\label{tab:atm_err}
\begin{center}
\begin{tabular}{lllllllllllll}
\hline\hline\noalign{\smallskip}

Ion & \multicolumn{2}{c}{$\Delta$\teff} & \multicolumn{2}{c}{$\Delta$log g}
& \multicolumn{2}{c}{$\Delta$\vt}
\\
    & AV\,303 & AV\,487~~~~~~ & AV\,303 & AV\,487~~~~~ & AV\,303 & AV\,487
\\
\hline
\\
C II~~~~~   & +0.10 & +0.21 & +0.01 & -0.18 & -0.01 & -0.01
\\
N II   & -0.09 & +0.22 & +0.09 & -0.21 & -0.08 & -0.01
\\
O II   & -0.30 & +0.23 & +0.16 & -0.24 & -0.09 & -0.03
\\
Mg II  & +0.14 & +0.14 & -0.07 & -0.17 & -0.01 & -0.07
\\
Si III & -0.30 & +0.27 & +0.18 & -0.23 & -0.15 & -0.06
\\
\noalign{\smallskip}   \hline
\noalign{\smallskip} \hline     
\end{tabular}
\end{center}
\end{table*}    
%------------------------------------------------------------------------------- 

%----------------------------------fastwind vs tlusty---------------------------

\section{Comparison of FASTWIND and TLUSTY analysis techniques}
\label{fast_tlusty}

The analysis presented in Sect.\ \ref{analysis} is based on static calculations
using the plane-parallel non-LTE model atmosphere code, {\sc tlusty}. Clearly
the neglect of the wind must be problematic for stars with relatively  large
luminosities and low gravities (see, for example Aller \cite{All56}; 
Kudritzki \& Puls \cite{Kud00} amongst others). Indeed features such as 
the Balmer H$\beta$ line in the spectra
of our current sample show clear evidence of the presence of a wind. Hence it
is useful to compare our results with those deduced from a unified code. 
In fact VLT/UVES spectra
of two of our targets (AV\,215 and SK\,191) have been previously analysed
by Trundle et al. (\cite{Tru04}), facilitating such a comparison.
The version of the code {\sc fastwind} used by Trundle et al. included some updates 
from that introduced by Santolaya-Rey et al. (\cite{San97}) for line blanketing 
and blocking and is briefly described in Herrero et al. (\cite{Her02}) 
and Repolust et al. (\cite{Rep04}).

\begin{table*}
\begin{center}
\caption[aat_vlt]
{Comparison of atmospheric parameters and abundance estimates for AV215 
and SK191}
\begin{tabular}{lrrrrrr}
\\
\hline
\noalign{\smallskip}
&             \multicolumn{3}{c}{AV215}    &  \multicolumn{3}{c}{SK191} \\
& TL/AAT & TL/VLT & FW/VLT & TL/AAT & TL/VLT & FW/VLT \\
\noalign{\smallskip}
\hline	     
Teff   &      26500  & 26500    &  27000   & 20500 & 20500 & 22500 \\
log g  &      3.00   & 2.95     &  2.90    & 2.45  & 2.45  & 2.55  \\
\vt     &      13     & 12       &  12      & 16    & 15    & 13    \\
\hline
\\
\ion{C}{ii}   &      -      & 7.00     & 6.91     & 7.00  & 6.88  & 6.89  \\
\\
\ion{N}{ii}   &      7.60   & 7.80     & 7.96     & 7.53  & 7.45  & 7.63  \\
\\
\ion{O}{ii}   &      8.00   & 7.98     & 7.97     & 8.11  & 8.20  & 8.20  \\
\\
\ion{Mg}{ii}  &      6.67   & $<$7.20  & $<$7.20  & 6.71  & 6.87  & 6.98  \\
\\
\ion{Si}{iii} &      6.87   & 6.90     & 7.10     & 6.64  & 6.70  & 6.75  \\
\ion{Si}{iv}  &      6.89   & 6.85     & 7.14     & 6.61  & 6.51  & 6.50  \\
\\
\noalign{\smallskip}
\hline
\label{tl_fw}
\end{tabular}
\end{center}
\end{table*}

Although it is possible to directly compare the results of
Trundle et al. (hereafter designated FW/VLT) with those found here
(designated TL/AAT), this has the disadvantage that both the 
observational data and theoretical methods differ between the 
two analyses. To make the comparison easier, we have therefore 
taken the VLT spectra and analysed them using our {\sc tlusty} grid
and the same procedures as discussed above. This analysis is designated
TL/VLT and the results for all three analyses in both stars are
summarized in Table \ref{tl_fw} and are discussed below.

We stress that the comparison discussed below does not attempt to 
determine which of the codes has a better description of the physical
processes relevant to the complicated atmospheres of B-type supergiants. 
Instead, it has the more limited objective of trying to quantify 
and understand the differences in the estimates for the atmospheric parameters 
and chemical composition obtained for such objects when using codes which 
incorporate different physical assumptions and processes.

\subsection{Observational uncertainties}

A comparison of the TL/AAT and TL/VLT analyses gives an insight into the
effects of observational uncertainties on the resulting model atmosphere
analysis. The VLT/UVES spectra discussed by Trundle et al.\ had S/N ratios
in the blue spectra region of 120 (AV\,215) and 170 (SK\,191), compared
with 90 and 125 respectively for the UCLES/AAT data. However this 
comparison is misleading as the VLT/UVES data had been re-binned to
0.2\AA, whilst the pixel size for the UCLES/AAT spectra was 0.09\AA.
When the latter is binned to a pixel size of 0.2\AA, the S/N ratios
of the two spectra become comparable, as is illustrated in Fig.\ 
\ref{vlt_aat}.

\begin{figure}[ht]
\includegraphics[angle=270,width=3.4in]{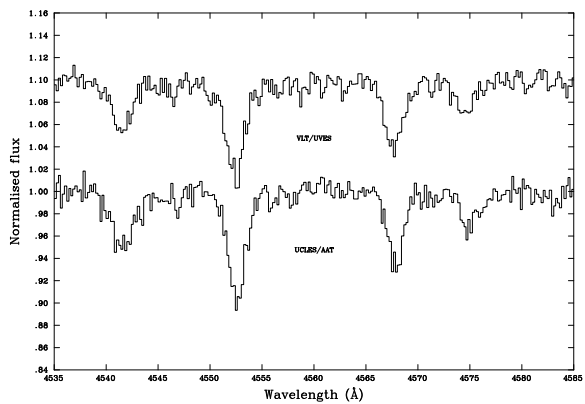}
\caption{
Comparison of AAT/UCLES and VLT/UVES spectra for the star AV\,215.
The spectra have both been rebinned to a pixel size of 0.2\AA\ and 
the VLT/UVES spectrum has been shifted by 0.1 in y-direction to aid
comparison. The spectral region includes the \ion{He}{ii} line at 
4541\AA\ and the \ion{Si}{iii} multiplet ar 4552, 4567 and 4575\AA}
\label{vlt_aat}
\end{figure}

The agreement between the two analyses is encouraging with very similar
atmospheric parameters being estimated. For example, the effective 
temperature estimates agree for both stars with differences in the
gravity and microturbulence estimates being at most 0.05 dex and
1\kms respectively. For the abundance estimates, in only three cases
is the discrepancy greater than 0.1 dex with the maximum discrepancy 
being 0.2 dex. For two cases (\ion{C}{ii} and \ion{Mg}{ii} in SK\,191) this 
reflects a difference in the equivalent width of the single line 
used to estimate the abundance. For \ion{N}{ii}, the discrepancy
arises both from differences in the equivalent width measurements and
in the sets of lines considered. Hence we conclude that for observational
data of this quality, the corresponding uncertainties are relatively small
with abundance estimates having a typical accuracy of $\pm$0.1 dex.

\subsection{Theoretical uncertainties}\label{th_unc}

A more complicated but potentially more interesting comparison is that which
adopts the same observational data but utilises different theoretical
approaches, viz. the TL/VLT and FW/VLT analyses. The effective temperature
estimates differ by 500K (for AV\,215) and 2000K (for SK\,191) with the {\sc
fastwind} estimates being higher. We have investigated these differences by
comparing the temperature structures in the two sets of models. Initially the
{\sc fastwind} code adopts a {\sc tlusty} temperature structure. This is then
altered by the considerations of sphericity and with the requirement that
continuity is obtained at the transition point between the photosphere and
stellar wind. The stronger the stellar wind included in a model at a 
particular temperature the further  into the stellar atmosphere this transition
occurs and once the wind becomes important {\sc fastwind} assumes an isothermal
atmosphere at a temperature appropriate to the given atmospheric parameters.
SK\,191  has a relatively strong wind for its spectral type and the requirement
for an isothermal atmosphere in the wind at a predefined temperature minimum
leads to the {\sc fastwind} models having lower temperatures in the atmospheric
region where the silicon lines are formed (by approximately 1\,000 K). Since
these Si lines are important temperature indicators in this stellar regime,
this naturally affects the effective temperature estimates derived from the two
codes. For AV\,215, the wind becomes important at a similar Rosseland optical
depth to that  of SK\,191. However the {\sc fastwind} temperature structure is
in close  agreement to that of the {\sc tlusty} model consistent with the
better  agreement in the effective temperatures estimates for this star.  The
latest version of {\sc fastwind} (Urbaneja \cite{Urb04}; Puls et al.
\cite{Pul05}) incorporates new techniques for delineating the  temperature
structure. These ensure thermal balance for the electrons (see Kub\'{a}t, Puls
\& Pauldrach, \cite{Kub99}), as an alternative to solving the radiative
equilibrium equation explicitly. Puls et al. show that the new temperature
structures calculated by {\sc fastwind} are in very good agreement with those
obtained by the unified model atmosphere codes {\sc cmfgen} (Hillier \& Miller
\cite{Hil98}) \& {\sc wm-basic} (Pauldrach et al.\,\cite{Pal01}). We have not yet
implemented this version, but it may lead to a more accurate and better 
understanding of how the wind influences the photospheric temperature 
structure.

\begin{figure}[ht]
\includegraphics[angle=0,width=3.4in]{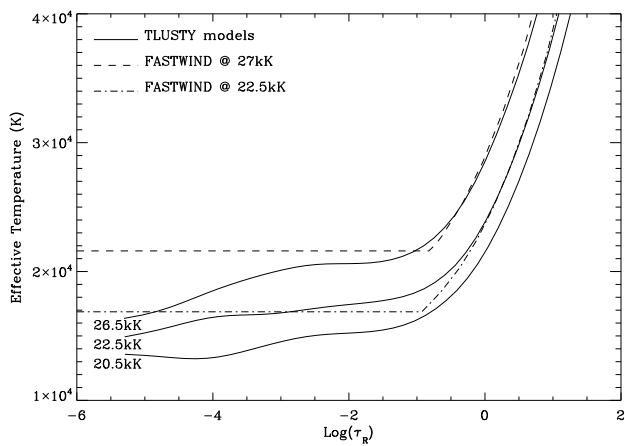}
\caption{Temperature structures of {\sc tlusty} and {\sc fastwind} atmospheric
models. The solid lines represent the {\sc tlusty} models with effective
temperatures of 26.5, 22.5 and 20.5 kK as marked on the figure. The first
represents the model parameters for AV\,215 and the latter two that for 
SK\,191. The
dashed and dashed-dotted line represent the equivalent {\sc fastwind} models 
for effective temperatures of 27.0 and 22.5 kK with the appropriate wind 
parameters for AV215 and
Sk191, respectively. All other parameters are consistent between
these models. Notice the Rosseland optical depth at which the wind becomes
important in the {\sc fastwind} models; identified by the isothermal behaviour
of the temperature structure. 
}
\label{temp_struc}
\end{figure}

The gravity estimates differ by 0.05 to 0.10 dex, although particularly
for SK\,191 this difference will reflect, at least in part, the different
estimates adopted for the effective temperature. The microturbulence estimates
are in relatively good agreement with the maximum difference  being 2\,\kms.
Given the different physical assumptions, model ions and numerical techniques
used in the two sets of calculations, the agreement  must be considered
encouraging.

It is also possible to compare the abundance estimates but it should
be noted that this will be complicated by differences in the adopted
atmospheric parameters. For AV\,215, the two sets of parameters were
effectively the same but for SK\,191, the effective temperature estimates
differed by 2\,000K (although the differences in temperature structure 
in the regions where the lines are formed may be smaller). Hence we 
will consider the two stars separately.

For AV\,215, the C, O and upper limit for Mg abundance estimates are in 
excellent agreement. By contrast, the N abundance estimates differ by 0.16 dex.
This is surprising as previous calculations (Becker and Butler \cite{Bec89})
have show that non-LTE effects in the \ion{N}{ii} are relatively  small
compared with, for example, those in \ion{C}{ii} (Sigut \cite{Sigut}). The
differences for \ion{Si}{iii} and \ion{Si}{iv} estimates are even larger and
range from 0.2 to 0.3 dex.

We have attempted to understand the cause of these differences as follows.  An
increase in the adopted mass-loss rate changes the density structure  leading
to a decrease in the upper photosphere and in the wind.  In addition the flux
in the Paschen continuum is decreased and  this causes the reduction in the
line strength observed in the \ion{Si}{iii} lines shown in Fig. \ref{si3}. By
contrast the \ion{Si}{iv} 4116\AA\, lines (see Fig. \ref{si4}) show no
significant change in their line strength, whilst the \ion{He}{ii} 4541\AA\,
line strength is actually increased. As the latter represent the
predominant ionization stage, this suggests a change in the ionisation
equilibrium, which could be obtained by changing either the temperature or 
density structures. Test calculations for an effective temperature of 27\,000K
show that the difference in the temperature structure (of upto 1\,000K) 
induced by increasing the wind from $10^{-7}$M$_{\odot}$yr$^{-1}$ to 
$1.35\times10^{-6}$M$_{\odot}$yr$^{-1}$ cannot account for the change  in the
Si line strengths, suggesting that it is variations in the density  structure
which might be more important. Without changing the gravity,  the only way to
alter the density structure is to change the  $\beta$ parameter which controls
the velocity field in the wind.  Increasing/decreasing the value of $\beta$
shifts a high mass-loss rate  model to higher/lower ionisation stages i.e
decreasing $\beta$ increases  the density. In the case of the {\sc fastwind}
model for AV215, lowering  the $\beta$\, value to 1.1 (and appropriately 
adjusting the
mass-loss rate so the  H$\alpha$ profile is adequately reproduced) decreases
the \ion{Si}{iv}  and increase the \ion{Si}{iii} line strengths, in such a way
that the  Si abundance estimate can be reduced by approximately 0.1 dex. The 
complex dependence of the Si line strengths on the temperature structure, 
microturbulence, and the wind parameters introduces some further uncertainty 
for the silicon abundance estimates found from a unified model atmosphere 
code, and this may be reflected in the larger uncertainties quoted by  Trundle
et al. than for the {\sc tlusty} analysis presented here.

\begin{figure}[ht]
\includegraphics[angle=0,width=3.4in]{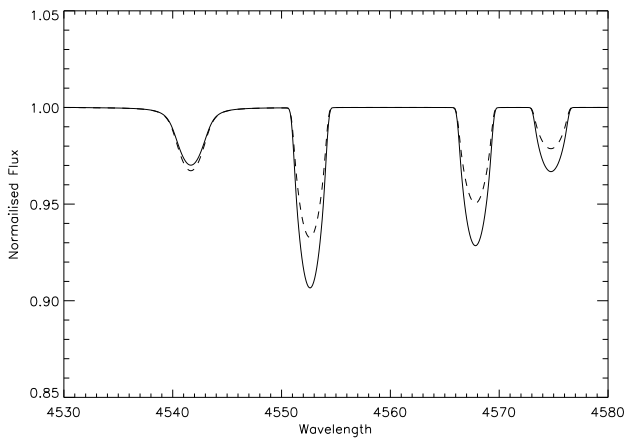}
\caption{Effect of stellar wind on the \ion{Si}{iii} multiplet at 4552-4575A. 
The spectra are for a {\sc fastwind} model at atmospheric parameters 
appropriate for AV215 (dashed line) and a model with the mass-loss rate 
reduced to 15\% of its original value (solid line) to simulate a 
plane-parallel model. 
Note the affect on the silicon lines due to the inclusion of the stellar 
wind. These theoretical spectra have been convolved with a projected rotational 
velocity of 91\kms, consistent with the value found for AV215 
by Trundle et al.  (\cite{Tru04}).
}
\label{si3}
\end{figure}

\begin{figure}[ht]
\includegraphics[angle=0,width=3.4in]{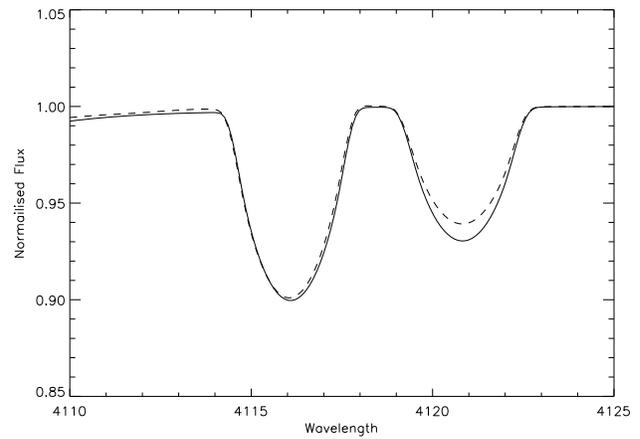}
\caption{Effect of stellar wind on the \ion{Si}{iv} line at 4416\AA. The
spectra are for the same models as shown in Fig. \ref{si3}. Note that for
this line, the wind appears to have little effect on the line strength. 
}
\label{si4}
\end{figure}

For SK\,191, the C, O and Si abundance estimates are in excellent agreement,
although given the different adopted effective temperatures, this may be
fortuitous. The Mg values differ by 0.12 dex but this is consistent with the
lower effective temperature used in the TL/VLT analysis.  Finally the N
estimates differ by 0.18 dex with the discrepancy being in the same  sense as
for AV\,215. 

We have investigated whether the nitrogen model ions in the two calculations 
could explain this discrepancy. We note that the \ion{N}{iii} model ion adopted 
in the {\sc tlusty} grid, split the gound state $^2$P term  into the two levels
with different J values. However we have carried out test calculations using a
simpler \ion{N}{iii} model ion with a single ground state term and find that
this has a negligible effect on the predicted  \ion{N}{ii} spectra. Hence we
do not believe that the adopted \ion{N}{iii} model ions are the cause
of the discrepancy. We have
also reformatted the {\sc tlusty} \ion{N}{ii} model ion so that it could be
used by the {\sc fastwind} code.
Test calculations with this new model ion at atmospheric conditions appropriate
for SK\,191 and AV\,215, showed that for the same physical conditions, the 
equivalent widths of singlet \ion{N}{ii} lines (viz. 3995, 4227, 4447 \AA) 
were increased by no more than 4\% whilst the triplets (viz. the 4630 \AA\ 
multiplet) differed by up to 40\%. The latter discrepancy does not have a 
large effect on the mean nitrogen abundance adopted from both analyses but 
is simply reflected in the larger standard deviations for the {\sc fastwind} 
analysis. 

In the case of SK\,191, the discrepancy in the nitrogen abundance may also 
reflect the difference in the atmospheric structures between the {\sc
fastwind}  and {\sc tlusty} models (see Fig.~\ref{temp_struc}) in the region of
formation  of the nitrogen lines. Additionally the N line strengths deduced
from {\sc fastwind} models decrease as the mass-loss rate is inceased due to
changes in the Paschen continuum flux. For example reducing the  mass-loss rate
for SK\,191 by 15\% in test  calculations with {\sc fastwind} had the effect of
increasing the equivalent  widths of the nitrogen lines by $\sim$ 25\%, while
for AV\,215 the  omission of the wind increases the nitrogen equivalent widths
by 38\%.  Hence we believe that the lower nitrogen abundances estimates deduced
from {\sc tlusty} calculations compared with those from the {\sc fastwind}
calculation are due  at least in part to the inclusion of the effect of the
stellar wind on the  photospheric lines in the latter.

Given the different physical assumptions, model ions etc, used in the two
analyses the agreement in the abundance estimates is surprisingly good. For
example, the mean of the {\it modulus} of the differences in the estimates is
only 0.10 dex. In turn this implies that even for these luminous Ia
supergiants, an analysis using static atmospheres, although not ideal, may
still be valid. This is important as in some cases particularly for faint
extragalactic objects, the observational material to constrain the wind
parameters may not be available. 
However our two supergiants are not amongst the most extreme 
Ia$^+$\,objects, where the effects observed here (such as filling in
of the lower Balmer lines) might become more significant for both the higher
Balmer lines and the metal line spectra. Additionally for such extreme 
objects the wind together with geometrical effects could influence the 
temperature structure. 

Although this comparison has been limited to only two supergiants, it has
yielded some important results. Firstly to reliably model the totality of 
the optical spectra, it is necessary to use a model that explicitly
includes the stellar wind. This is, for example, important for the 
lower members of the Balmer series. However by using features which are 
not significantly affected by the stellar wind (such as higher members 
of the Balmer series) when determining the stellar parameters and 
chemical compositon, an analysis using static models may be appropriate. 
Indeed the complexity of an analysis with {\sc fastwind} or other unified 
models (as illustrated by our discussion of the Si spectrum) can introduce
additional uncertainties. However probably the most important and 
encouraging result from this comparison is that both approaches appear to 
lead to reliable estimates for both the atmospheric parameters and 
chemical compositions.

%--------------------------------------------------------- tab: mean ------- 
\begin{table*}
\begin{center}
\caption{Mean absolute  abundances for the program stars. Also
listed are results for analyses of other targets in the SMC, viz. 
B-type supergiants, Trundle et al. (\cite{Tru04},TLPD); 
A-type supergiants, Venn (\cite{Ven99}, Venn \& Pryzbilla (\cite{Ven03a});
NGC 330 giants, Lennon et al. (\cite{Len03}); AV\,304, Hunter et al.
(\cite{Hun04}); \ion{H}{ii} regions Kurt et al. (\cite{Kur99}). Note a 
correction has been applied to the C abundance in the TLPD analysis, 
due to the problematic 4267\ \AA line. If such a correction was applied 
to our estimate excellant agreement is found (see discussion in 
Sect.~\ref{Discussion}).}
\label{tab:mean}
\begin{tabular}{lllllllllllll}
\hline\hline\noalign{\smallskip}
Objects  & \multicolumn{3}{c}{AB-type supergiants} & \multicolumn{2}{c}{B-type stars}
& \ion{H}{ii} 
\\
Analysis &  this work & TLPD & A-type & NGC330 & AV304 & regions
\\
\hline\\
C        & 7.06$\pm$0.12  & 7.30$\pm$0.09 &           -- 
         & 7.26$\pm$0.15  & 7.36$\pm$0.12 & 7.53$\pm$0.06
\\
N        & 7.42$\pm$0.15  & 7.67$\pm$0.27 & 7.52$\pm$0.10 
         & 7.51$\pm$0.18  & 6.55$\pm$0.18 & 6.59$\pm$0.08
\\
O        & 8.09$\pm$0.12  & 8.15$\pm$0.07 & 8.14$\pm$0.06 
         & 7.98$\pm$0.15  & 8.12$\pm$0.10 & 8.05$\pm$0.05
\\
Mg       & 6.70$\pm$0.10  & 6.78$\pm$0.16 & 6.83$\pm$0.08 
         & 6.59$\pm$0.14  & 6.77$\pm$0.16 &           --
\\
Si       & 6.85$\pm$0.12  & 6.74$\pm$0.11 & 6.92$\pm$0.15 
         & 6.58$\pm$0.32  & 6.74$\pm$0.19 & 6.70$\pm$0.20 
\\
\\
\noalign{\smallskip}   \hline
\noalign{\smallskip} \hline     
\end{tabular}
\end{center}
\end{table*}    
%------------------------------------------------------------------------------- 

%------------------------------------------------------------ sec: discussion ----
\section{Discussion}       
\label{Discussion}
\subsection{Chemical compositions}

In Table \ref{tab:mean}, we summarize the mean abundances obtained from our
supergiant sample. The errors include the estimated standard deviations  of
these means, which should allow for {\em random} errors in the observational
data and adopted atmospheric parameters. Additionally {\em systematic} errors
in the effective temperature and gravity, will to some extent manifest 
themselves as random errors in the abundance estimates due to their different
effects on the abundance estimates in stars with various atmospheric
parameters (see Table \ref{tab:atm_err}). By contrast a systematic 
mis-estimation of the microturbulence (see Sect. \ref{vt} for details) would 
lead to systematic errors in the abundances. We have estimated these 
systematic errors to be of the order of 0.1 dex and have included them in
quadrature with the random errors. Note that these error estimates do not
include uncertainties due to, for example, limitations in the adopted model
ions or in the physical assumptions adopted in the calculations. Also included
in Table \ref{tab:mean} are results for analyses of other targets in the  SMC,
viz. B-type supergiants, Trundle et al. (\cite{Tru04},TLPD);  A-type
supergiants, Venn (\cite{Ven99}), Venn \& Pryzbilla (\cite{Ven03a}); NGC 330
giants, Lennon et al. (\cite{Len03}); AV\,304, Hunter et al. (\cite{Hun04}) and
\ion{H}{ii} regions, Kurt et al. (\cite{Kur99}).

Trundle et al. (\cite{Tru04}) presented non-LTE analyses of the spectra
of eight SMC supergiants. Our analysis of nine supergiants
(two of which are in common with the sample of Trundle et al.) 
effectively doubles the sample size, although our observational dataset
and methodology limits our analysis to those features that can be
modelled by a static photosphere. As can be seen from Table \ref{tab:mean},
the current results are in good agreement with those of Trundle et al.
for the elements O, Mg, Si. There would appear to be a discrepancy
of 0.24 dex in the mean C abundances found in the two analyses. However 
the value of Trundle et al. was primarily based on the \ion{C}{ii} doublet at 
4267\AA, which has been found to give systematically lower abundances 
estimates than other features (Eber \& Butler \cite{Ebe88}) and leads to
a mean abundance of 6.96 dex. Trundle et al. then corrected this value
using the methodology discussed by Lennon et al. (\cite{Len03}) to obtain
the value quoted in Table \ref{tab:mean}. Our C abundance estimates are
based on both the \ion{C}{ii} feature at 4267\AA\ and the lines at 3919 and
3921\AA. However if we correct our results for the former using the
same methodolgy, our mean C abundance becomes 7.32$\pm$0.15 dex,
which is in excellent agreement with that obtained by Trundle et al.
and indeed with other analyses of SMC objects.

For nitrogen, our mean abundance is 0.25 dex lower than that of 
Trundle et al., although it is in reasonable agreement with that
of other evolved SMC targets. Part of this difference might be
due to the possible systematic difference of approximately 0.15 dex found in
the N abundance estimates deduced from the {\sc tlusty} and {\sc fastwind}
analyses discussed in Sect. \ref{fast_tlusty}. However as discussed by
Trundle et al., B-type supergiants show a wide range of nitrogen 
enhancements (compared to unevolved objects) and hence at least part of 
this difference may reflect real variations in the mean nitrogen abundances
for the two samples. 

In Table \ref{tab:smc_sup}, we list the mean abundances  found for 
B-type supergiants when the current results
are combined with those of Trundle et al. For elements, N, O, Mg, Si,
the estimates have been simply averaged, whilst for C, our estimates have
been corrected as discussed above. The error estimates assume that
the errors follow a normal distribution (i.e. they are the standard deviation
of the individual estimates divided by the square root of the number of
measurements). As such they will not include any systematic errors due
to, for example, limitations in the physical assumptions. Such errors are
very difficult to assess but we note that in Sect.\,\ref{th_unc} the adoption
of different theoretical approaches yielded differences in the abundance 
estimates of typically 0.1 dex. These values in Table \ref{tab:smc_sup} 
represent our best estimates for the mean abundances of 
B-type supergiants in the SMC.

%--------------------------------------------------------- tab: mean ------- 
\begin{table}
\begin{center}
\caption{Mean absolute abundances for the B-type supergiants analysed here 
and by Trundle et al. (\cite{Tru04},TLPD). These values represent 
our best estimates for the mean abundances of B-type supergiants in the SMC.}
\label{tab:smc_sup}
\begin{tabular}{lllllllllllll}
\hline\hline\noalign{\smallskip}
Element   & Abundance 
\\
\hline
\\
C        & 7.30$\pm$0.04 
\\
N        & 7.55$\pm$0.08 
\\
O        & 8.11$\pm$0.04  
\\
Mg       & 6.75$\pm$0.03  
\\
Si       & 6.80$\pm$0.04  
\\
\noalign{\smallskip}   \hline
\noalign{\smallskip} \hline     
\end{tabular}
\end{center}
\end{table}    
%------------------------------------------------------------------------------- 

The elements, O, Mg and Si, are unlikely to have been affected by
the mixing of nucleosynthetic material to the surface and the agreement
between these mean supergiant abundances and those for AV\,304
is excellent with differences of 0.06 dex
or less. This provides indirect evidence that the methods used for
the supergiant spectra are reliable. For C, the supergiants's
estimate is smaller than that for AV\,304 by 0.06 dex, with N being typically
enhanced by 1.0 dex. As discussed by Trundle et al.\ this is consistent
with processed material having been mixed to the surface and with the
predictions of Maeder \& Meynet (\cite{Mae01}). Unfortunately Maeder \& Meynet
only tabulate abundances ratios, whilst their initial abundances are
one fifth solar, corresponding to a N abundance of 7.3 dex, 
significantly greater than that found in AV\,304 and SMC \ion{H}{ii} regions.
Hence it is not straightforward to directly compare our absolute abundances
with existing stellar evolutionary calculations. However it is interesting to
note that stars, such as AV\,78, have currently N abundances that are larger
than the sum of their presumed initial C and N abundances. Thus if all the
additional nitrogen has been formed via hydrogen burning some depletion
of oxygen would also be required. Indeed besides a very high N abundance,
AV\,78 also exhibits relatively low C and O abundance estimates.

As discussed by, for example, Trundle et al. (\cite{Tru04}), current stellar
evolutionary calculations require large initial stellar rotational velocities
in order to obtain significant N enhancements at the surface of B-type
supergiants. Additionally, it is predicted that these objects will still be
rotating relatively quickly; for example, the models of Maeder and Maynet
(\cite{Mae01}) with initial rotational velocities of 300\,\kms and masses of
20 to 60 solar masses have predicted rotational velocities of 100 to 200\,\kms
for effective temeparture appropriate to early B-type supergiants. However
Trundle et al. observed metal line widths, which would imply projected
rotational velocities of 50-90\,\kms and even allowing for projection effects,
these would appear incompatible with the predictions. We find similar 
inconsistencies with evolutionary models with the FWHM listed in 
Table \ref{tab:obs} again implying projected rotational velocities of 
less than 100\,\kms.

The discrepancy may be greater than this simple comparison implies, 
as it assumes that the observed line widths
are dominated by rotation. As discussed by Howarth et al. (\cite{How97}),
the lack of any B-type supergiants with narrow lines implies that another
broadening mechanism, as well as rotation, must be present. This was 
confirmed by Ryans et al. (\cite{Rya02}), who analysed very high quality
spectra of Galactic B-type supergiants to distinquish between the effects
of rotation and turbulence. They found that turbulence was the dominant
mechanism with estimates of projected rotational velocities being typically
10-20\,\kms. For the current SMC dataset, it was found that Gaussian profiles
gave a good fit to the observed metal line spectra again implying that
rotation was not the dominant mechanism. Additionally, analyses of the spectra
of O-type main sequence stars normally indicate N enrichment in their
atmospheres. For example Heap et al. (\cite{Hea04}) found that for seventeen
O-type stars, fourteen of them exhibited N enrichment. Hence it is clear that
the mixing process operates at an early evolutionary stage and either that 
it occurs at 
relatively small rotational velocities or that the velocity breaking between 
B-type supergiants and their O-type precursors is greater than predicted by 
current evolutionary models. 

Venn (\cite{Ven99}) and Venn and Przybilla (\cite{Ven03a}) have estimated N 
abundances for SMC A-type supergiants. As can be seen from Table 
\ref{tab:mean},
their mean value is in excellent agreement with the value found by combining 
our two samples (see Table \ref{tab:smc_sup}). In turn this implies that 
little further enrichment occurs as B-type supergiants evolve into
A-type supergiants, which is consistent with the predictions of evolutionary
calculations (see, for example, Maeder and Maynet \cite{Mae01})

\section{Conclusions}

We have presented a grid of 12\,000 non-LTE models covering the range of
atmospheric parameters appropriate to B-type stars. These should allow 
efficient and reliable analysis of such objects in environments with  different
metallicities. In this paper, the grid has been used to analyse the spectra of
9 SMC B-type supergiants to obtain atmospheric parameters and chemical
compositions. The principal results of this work can be summarized as follows:

\begin{enumerate}

\item  The abundance estimates for O, Mg and Si are in excellent agreement with
those deduced from unevolved stars and \ion{H}{ii} regions and provide
indirect support for the reliability of the methods. The high N abundances
found in other evolved objects (see, for example, Korn et al. \cite{Kor02};
Trundle et al. \cite{Tru04}, Venn \cite{Ven99}) are found here and for 
the most extreme cases imply that both the CN and ON cycles must have been 
operating. 
\item For two stars we have compared the results obtained using our {\sc
tlusty} grid with those deduced using the unified code {\sc fastwind}. Even for
these luminous Ia supergiants the agreement is excellent with 
discrepancies being of a similar magnitude to the observational uncertainties. 
\item Our estimates of the upper limits for the projected rotational velocities
appear inconsistent with those required by current evolutionary models showing
significant enhancements of N at the stellar surfaces. This clearly warrants
further investigation into the actual projected rotational velocity of 
B-type SMC supergiants (rather than just upper limits)
and in a companion paper we analyse using the methods discussed by Ryans 
et al. (\cite{Rya02}) both the dataset 
presented here and that of Trundle et al. (\cite{Tru04}) 
. 
\end{enumerate}

\begin{acknowledgements}

We are grateful to the staff of the Anglo-Australian Telescope for their
assistance. RSIR and CAP acknowledge financial support from the PPARC
and from NASA (ADP02-0032-0106 and LTSA 02-0017-0093) respectively.
The non-LTE calculations were facilitated by the use of software from the
Condor Project (see http://www.cs.wisc.edu/condor/). We thank Joachim Puls
for his useful comments on a draft of this paper and Robert Rolleston 
for help and advice.

\end{acknowledgements}

\end{document}